\documentclass[aps,prd,superscriptaddress,long,notitlepage,balancelastpage,nofootinbib,floatfix,twocolumn]{revtex4-1}
\pdfoutput=1
\usepackage{amsmath,mathtools,amssymb,amsthm,amsxtra,overpic,bbm,epsfig,subfigure,url,bm}
\usepackage{hyperref}
\usepackage{mathrsfs}
\usepackage{color,xcolor}
\usepackage{comment}
\usepackage{float}
\usepackage{enumitem}
\usepackage{slashed}
\usepackage{multirow}
\usepackage{appendix}
\DeclareUnicodeCharacter{03BC}{\ensuremath{\mu}}
\DeclareUnicodeCharacter{2032}{\ensuremath{'}}
\definecolor{nicered}{rgb}{0.5,0.,0.}
\definecolor{nicegreen}{rgb}{0.,0.5,0.}
\definecolor{niceblue}{rgb}{0.,0.,0.5}
\hypersetup{
	colorlinks=true,
	linkcolor=black,
	filecolor=nicegreen,      
	urlcolor=niceblue,
	citecolor=nicered,
}

\makeatletter
\newcommand*{\balancecolsandclearpage}{%
	\close@column@grid
	\cleardoublepage
	\twocolumngrid
}
\makeatother
\begin{document}
%%%%%%%%%%%%%%%%%%%%%%%%%%%%%%

\title{\vspace{1cm} \large 
Non-Standard Interactions of Supernova Neutrinos\\ \vspace{0.02in} and  Mass Ordering Ambiguity at DUNE
}

%%%%%%%%%%%%%%%%%%%%%%%%%%%%%%    
\author{\bf Sudip Jana}
\email[E-mail:]{sudip.jana@mpi-hd.mpg.de}
\affiliation{Max-Planck-Institut f{\"u}r Kernphysik, Saupfercheckweg 1, 69117 Heidelberg, Germany}
%%%%%%%%%%%%%%%%%%%%%%%%%%%%%%

%%%%%%%%%%%%%%%%%%%%%%%%%%%%%%    
\author{\bf Yago Porto}
\email[E-mail:]{yago.porto@ufabc.edu.br}
\affiliation{Centro de Ciências Naturais e Humanas, Universidade Federal do ABC, 09210-170, Santo André, SP, Brazil}
%%%%%%%%%%%%%%%%%%%%%%%%%%%%%%

%%%%%%%%%%%%%%%%%%%%%%%%%%%%%%%%%%%%%%%%%%%
%%%%%%%%%%%%%%%%%%%%%%%%%%%%%%%%%%%%%%%%%%%
\begin{abstract}
We show that non-standard neutrino interactions (NSI) can notably modify the pattern of resonant flavor conversion of neutrinos within supernovae and significantly impact the neutronization burst signal in forthcoming experiments such as the Deep Underground Neutrino Experiment (DUNE). The presence of NSI can invert the energy levels of neutrino matter eigenstates and even induce a new resonance in the inner parts close to the proto-neutron star. We demonstrate how DUNE can use these new configurations of energy levels to have sensitivity to NSIs down to $\mathcal{O}(0.1)$. We also elucidate how the effect may result in a puzzling confusion of normal and inverted mass orderings by highlighting the emergence or vanishing of the neutronization peak, which distinguishes between the two mass orderings. Potential implications are analyzed thoroughly.
%%%%%%%%%%%%%%%%%%%%%%%%%%%%%%%%%%%%%%%%%%%
%%%%%%%%%%%%%%%%%%%%%%%%%%%%%%%%%%%%%%%%%%%
\noindent 
\end{abstract}
\maketitle
%%%%%%%%%%%%%%%%%%%%%%%%%%%%%%%%%%%%%%%%%%%
%%%%%%%%%%%%%%%%%%%%%%%%%%%%%%%%%%%%%%%%%%%
\textbf{\emph{Introduction}.--}
In recent decades, extensive efforts and data from solar, atmospheric, reactor, and accelerator neutrino experiments have provided robust evidence for neutrino oscillations, indicating the presence of neutrino masses and mixing. However, the origin of these phenomena remains unestablished, leaving room for potential new physics, particularly in the form of non-standard neutrino interactions (NSIs). First introduced by Wolfenstein in 1978 \cite{Wolfenstein:1977ue}, NSIs have been the subject of intense scrutiny since then, providing an avenue for exploring new aspects of neutrino physics. These interactions involve higher-dimensional operators with neutrinos and matter, as represented by the equations: 
\begin{equation}
\begin{aligned}
& \mathcal{L}_{\mathrm{NC}}=-2 \sqrt{2} G_F \sum_{f, P, \alpha, \beta} \varepsilon_{\alpha \beta}^{f, P}\left(\bar{\nu}_\alpha \gamma^\mu P_L \nu_\beta\right)\left(\bar{f} \gamma_\mu P f\right) \\
& \mathcal{L}_{\mathrm{CC}}=-2 \sqrt{2} G_F \sum_{f, P, \alpha, \beta} \varepsilon_{\alpha \beta}^{f, P}\left(\bar{\nu}_\alpha \gamma^\mu P_L \ell_\beta\right)\left(\bar{f} \gamma_\mu P f^{\prime}\right)
\end{aligned}
\end{equation}
where $\varepsilon$ represents the strength of NSI relative to the weak scale, $P \in {P_L, P_R}$ indicates chirality projection operators and the sum is over matter fermions $f, f^{\prime} \in {e, u, d}$. 
Such NSIs modify the matter potential experienced by neutrinos, adding substantial intricacy to the determination of neutrino oscillation parameters. For instance, the existence of NSI introduces ambiguity in the determination of $\theta_{12}$ from the solar neutrino data \cite{Miranda:2004nb}. Furthermore, NSI effects have been shown to alleviate the tension between solar and KamLAND data, as it flattens the solar neutrino spectrum at high energies ($>3$ MeV) and generates larger day-night asymmetry \cite{Maltoni:2015kca}. In this study, we advocate for the utilization of supernova neutrinos to investigate the influence of these NSI on the precise determination of neutrino oscillation parameters.

\begin{figure}[htb!]
\centering
  \includegraphics[width=0.5\textwidth]{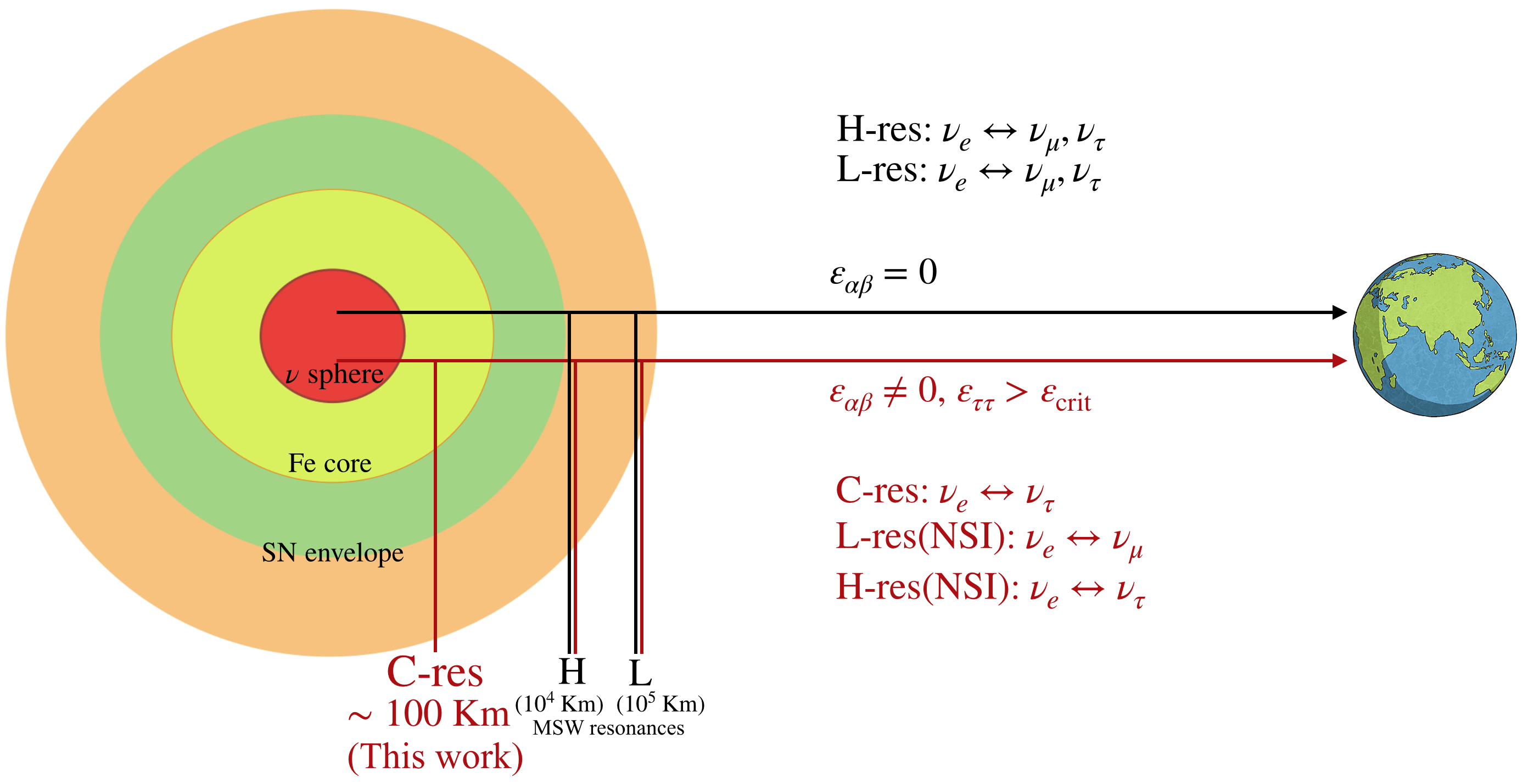}~~
  \caption{A simplified picture of flavor conversions of supernova neutrinos in presence of NSI.}
\label{illustartion}
\end{figure}

During a supernova (SN) explosion, as the progenitor's core collapses to form a neutron star, about $10^{53}$ erg of gravitational binding energy is released in the form of neutrinos \cite{Colgate:1966ax, Arnett:1966, Bethe:1985sox, Wilson:1985, Janka:2017vlw}. Originating in the SN core with energies in the tens of MeV, these neutrinos travel through the stellar mantle and envelope, undergoing modifications in mixing and oscillations through neutrino-matter interactions\footnote{Matter effects can also influence the evolution of high-energy astrophysical neutrinos with energies greater than 100 TeV \cite{Dev:2023znd}.} \cite{Wolfenstein:1977ue, Mikheyev:1985zog, Mikheev:1986wj, Mikheev:1986if}. The consequences of neutrino propagation in the SN-dense environment offer significant potential for probing new physics, and these effects are observable by flavor-sensitive Earth-based detectors \cite{Valle:1987gv, Nunokawa:1997ct, Nunokawa:1998vh, Esteban-Pretel:2007zkv, deGouvea:2019goq, Tang:2020pkp, Jana:2022tsa, Jana:2023ufy, dosSantos:2023skk, Bendahman:2023hjj}. 
%%%%%%%%%
A galactic SN, a rare event in the Milky Way occurring once or twice per century, was last observed in 1987 from the Large Magellanic Cloud, 51 Kpc away \cite{Tammann:1994ev, Kamiokande-II:1987idp, Bionta:1987qt, Alekseev:1988gp}. Neutrino detectors recorded about 25 $\bar{\nu}_e$ events during that event, significantly advancing our understanding of core-collapse processes and neutrino emission \cite{Olsen:2021uvt, Li:2023ulf, DedinNeto:2023hhp, Fiorillo:2023frv}. Future detectors, especially the Deep Underground Neutrino Experiment (DUNE), are anticipated to capture hundreds of thousands of events in various channels during a galactic supernova \cite{IceCube:2011cwc, DarkSide20k:2020ymr, Hyper-Kamiokande:2021frf, KM3NeT:2021moe, JUNO:2023dnp, DUNE:2020zfm}, with DUNE's unique capability to detect a clean $\nu_e$ signal in the critical first 40 ms after core bounce, known as the neutronization burst phase \cite{DUNE:2020zfm, Cuesta:2023nnt, Zhu:2018rwc}. In this article, we investigate the features of neutrino emission in the neutronization burst phase, elucidate the anticipated resonant flavor conversion within the stellar envelope triggered by NSI [cf.Fig~\ref{illustartion}] and discuss the crucial role of DUNE in exploring new physics impacting supernova neutrinos. 
%%%%%%%%%%%%%%%%%%%%%%%%%%%%%%%%%%%%%%%%%%%
%%%%%%%%%%%%%%%%%%%%%%%%%%%%%%%%%%%%%%%%%%%

\vspace{0.1in}

%%%%%%%%%%%%%%%%%%%%%%%%%%%%%%%%%%%%%%%%%%%
%%%%%%%%%%%%%%%%%%%%%%%%%%%%%%%%%%%%%%%%%%%
\textbf{\emph{Dynamics of Resonant Flavor Conversions}.--}
%%%%%%%%%%%%%%%%%%%%%%%%%%%%%%%%%%%%%%%%%%%
%%%%%%%%%%%%%%%%%%%%%%%%%%%%%%%%%%%%%%%%%%%
\begin{figure*}[!tb]
\centering
  \includegraphics[width=0.9\textwidth]{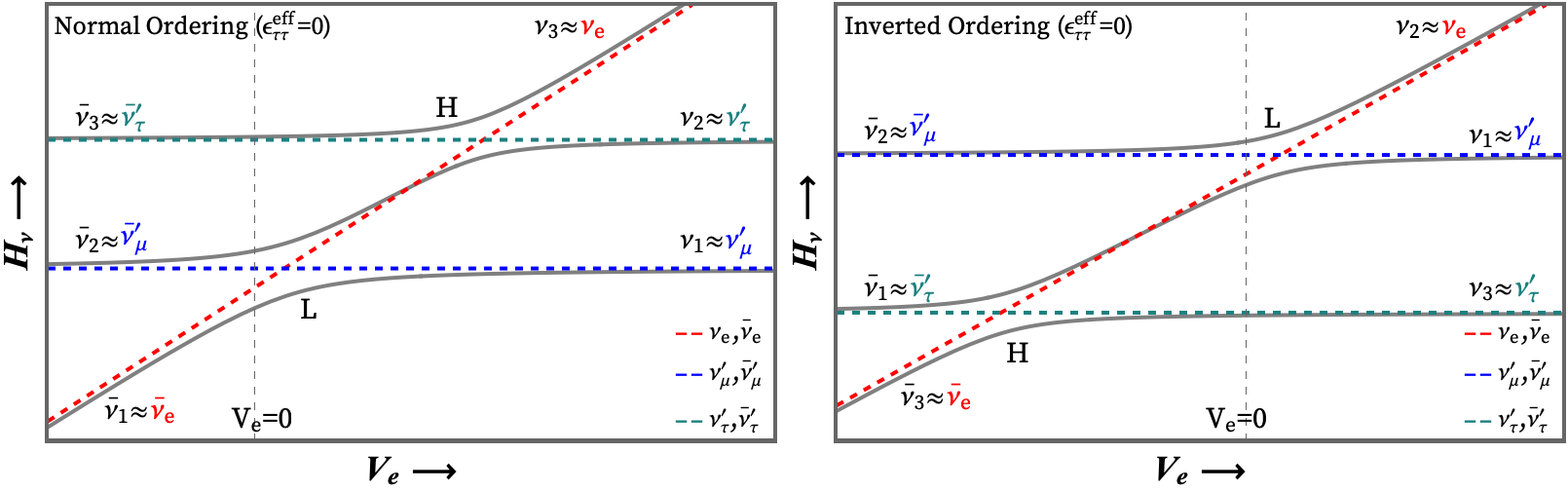}~~
  \caption{Configuration of energy levels for neutrinos and antineutrinos for NO (left) and IO (right). For each panel, neutrino lines are shown for positive $V_e$, while antineutrinos are plotted with negative $V_e$. The location of the MSW resonances is indicated by the letters $L$ and $H$. Solid lines represent the effective matter eigenstates, while the dashed lines follow the track of flavor states. Matter eigenstates mix with different flavors at different locations; this is shown by the flavor tags in red, blue, and green colors.}
\label{E-levels}
\end{figure*}
Understanding the flavor composition of SN neutrinos reaching the Earth during the neutronization burst phase requires monitoring how the initial fluxes generated in the core evolve while traveling outward within the star (see Appendix for details). In the dense environment where neutrinos propagate inside the star, vacuum oscillations are suppressed. However, the density variation in the way out of the SN instigates a flavor evolution dictated by a time (or space) dependent Hamiltonian that can trigger Mikheyev-Smirnov-Wolfenstein (MSW) resonant flavor conversion in specific layers of the matter profile \cite{Wolfenstein:1977ue, Mikheyev:1985zog, Mikheev:1986wj, Mikheev:1986if, Mikheev:1987jp}.

Assuming that neutrinos propagate radially outwards, the evolution equation for the flavor state $\nu=(\nu_e,\nu_\mu,\nu_\tau)^T$ is
\begin{equation}
    i \frac{d}{dr} \nu = \mathcal{H}  \nu.
\end{equation}
$\mathcal{H}$ is the Hamiltonian in the flavor basis given by
\begin{equation} \label{H}
    \mathcal{H}=\frac{1}{2 E} U\left(\begin{array}{ccc}
0 & 0 & 0 \\
0 & \Delta m_{21}^2 & 0 \\
0 & 0 & \Delta m_{31}^2
\end{array}\right) U^{\dagger}+V_{e} \left(\begin{array}{ccc}
1 & 0 & 0 \\
0 & 0 & 0 \\
0 & 0 & 0
\end{array}\right),
\end{equation}
where $E$ and  $U$ denote the neutrino energy and the Pontecorvo-Maki-Nakagawa-Sakata (PMNS) matrix. 
The matter potential due to charged current interactions can be expressed as 
\begin{equation}
    V_e=\sqrt{2}G_F \frac{\rho}{m_N} Y_e, 
\end{equation}
where $Y_e$ is the electron number fraction and $\rho$ is the matter density (see Fig.~\ref{profile} in the Appendix for the profile of $\rho$ and $Y_e$). The antineutrino Hamiltonian is similar to $\mathcal{H}$, with the only difference being that the matter potential inverts sign $\bar{V}_e=-V_e$. 

In dense matter, $\rho > 10^5$ $\text{g/cm}^3$, $V_e$ is much bigger than all other matrix elements in Eq.~\ref{H}, suppressing mixing and making $\nu_e$ the heaviest effective eigenstate in matter, $\nu_e \approx \nu_3^m$ for normal ordering (NO) and $\nu_e \approx \nu_2^m$ for inverted ordering (IO), the superscript $m$ denotes an effective eigenstate in matter. In a vacuum, however, $\nu_e$ is mostly associated with $\nu_1$. Therefore, as matter density decreases,  so does $V_e$, and $\nu_e$ will transit from $\nu_3^m$($\nu_2^m)$ to $\nu_1^m$. In doing so, it has to cross energy levels twice for NO ($\nu_3^m \rightarrow \nu_2^m \rightarrow \nu_1^m$) and once for IO ($\nu_2^m \rightarrow \nu_1^m$). The level crossing diagram is shown in Fig.~\ref{E-levels}. The red dashed line follows the evolution of the electron flavor, $\nu_e$ on the right of $V_e=0$ and $\bar {\nu}_e$ on the left. Note that $\bar {\nu}_e \approx \bar{\nu}_1^m$ in the inner regions, and there is no level crossing for antineutrinos in NO, while there is one level crossing in IO: $\bar {\nu}_3^m \rightarrow \bar {\nu}_1^m$. While $\nu_e$ and $\bar{\nu}_e$ are effective eigenstates in the SN environment, $\nu_\mu$, $\nu_\tau$, and their antiparticles are not effective eigenstates due to near maximal vacuum mixing. In this situation, it is convenient to diagonalize the $\nu_\mu-\nu_\tau$ subspace and work with the effective eigenstates $\nu'_\mu$, $\nu'_\tau$, and their antiparticles. The primed states have constant energy levels as a function of $V_e$, as shown in Fig.~\ref{E-levels}.

The crossing $\nu_e \leftrightarrow \nu'_\tau$ is called $H-$resonance and occurs when 
\begin{equation} \label{H-res}
    V_e(\rho_H) \approx \frac{\Delta m_{3 1}^2}{2 E} \cos \theta_{1 3},
\end{equation}
which corresponds to higher densities compared to the $L-$resonance, $\nu_e \leftrightarrow \nu'_\mu$, that happens when
\begin{equation} \label{L-res}
    V_e(\rho_L) \approx \frac{\Delta m_{2 1}^2}{2 E} \cos \theta_{1 2}.
\end{equation}
Note that Eq.~\ref{H-res} is satisfied for neutrinos in NO and for antineutrinos in IO ($\Delta m^2_{31}=-|\Delta m^2_{31}|$ and $V_e$ is negative) while Eq.~\ref{L-res} is satisfied only for neutrinos in both orderings. Using the profile in Fig.~\ref{profile}, and oscillation parameters from \cite{Esteban:2018ppq}, the $H-$resonance happens at $r\sim 10^4$ Km ($\rho_H$ in the range $10^3 - 10^4$ $\text{g/cm}^3$) and the $L-$resonance at $r \sim 10^5$ Km ($\rho_L \sim 1 - 10$ $\text{g/cm}^3$). Efficient resonant conversion only happens if neutrinos cross resonance layers adiabatically \cite{Dighe:1999bi}. We will assume perfect adiabaticity of the $H-$ and $L-$resonances so that transitions between distinct energy eigenstates are negligible. Moreover, the produced flux during the neutronization burst phase is mostly $\nu_e \approx \nu_3^m(\nu_2^m)$ for NO (IO). Due to adiabaticity, the initial eigenstates are preserved and reach vacuum as $\nu_3(\nu_2)$. Consequently, for NO only $|U_{e3}|^2 \approx 0.02$ of the initial $\nu_e$ flux, $F_{e}^i$, survives and reaches Earth, and some amount of $\nu_e$, ($1-|U_{e3}|^2) F_{ x}^i \approx 0.98 F_{ x}^i$, comes from the conversion of non-electron flavors, $F_{\mu}^i=F_{\tau}^i=F_{x}^i$:
\begin{equation} \label{FeNO}
    F_e^{N O} = |U_{e3}|^2 F_{ e}^i+(1-|U_{e3}|^2) F_{ x}^i \approx 0.02 F_{ e}^i+0.98 F_{ x}^i.
\end{equation}
Similarly for IO,
\begin{equation} \label{FeIO}
    F_e^{I O} = |U_{e2}|^2 F_{ e}^i+(1-|U_{e2}|^2) F_{ x}^i \approx 0.3 F_{ e}^i+0.7 F_{ x}^i.
\end{equation}

Due to $F_{e}^i$ being approximately ten times $F_{x}^i$ at the neutronization peak, the $\nu_e$ flux is greater in the Inverted Ordering (IO) than in the Normal Ordering (NO) during the initial 20 milliseconds of neutrino emission. This disparity enhances the visibility of the peak in neutrino detectors for the IO scenario. In the following sections, we will analyze signals derived from flux equations (\ref{FeNO}) and (\ref{FeIO}) in DUNE and discuss the presence or absence of the peak as a discriminative criterion between the two mass orderings. 

\vspace{0.1 in}
\begin{figure*}[htb!]
\centering
  \includegraphics[width=0.9\textwidth]{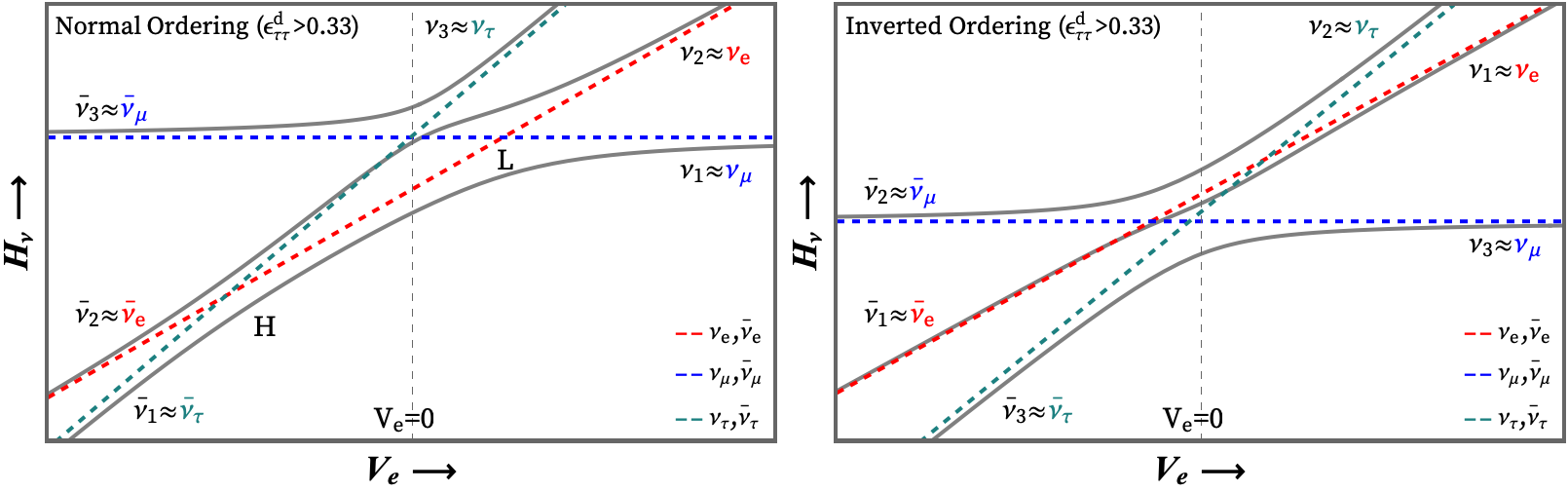}~~
  \caption{Configuration of energy levels for neutrinos and antineutrinos for NO (left) and IO (right)  with the inclusion of $\varepsilon_{\tau \tau}^{d}>0.33$. See text for details.}
\label{E-levels-NSI}
\end{figure*}

%%%%%%%%%%%%%%%%%%%%%%%%%%%%%%%%%%%%%%%%%%%
%%%%%%%%%%%%%%%%%%%%%%%%%%%%%%%%%%%%%%%%%%%
\textbf{\emph{Effect of NSI on the Energy Levels}.--}\label{sec-NSI}
%%%%%%%%%%%%%%%%%%%%%%%%%%%%%%%%%%%%%%%%%%%
%%%%%%%%%%%%%%%%%%%%%%%%%%%%%%%%%%%%%%%%%%%
The introduction of NSIs has a pronounced impact on the configuration of energy levels and the flavor composition of SN neutrinos detected on Earth. To demonstrate this effect, an additional energy term involving NSI parameters must be incorporated into the Hamiltonian mentioned in Eq.~\ref{H}. Our focus here is on NSI specifically influencing $\nu_\tau$, potentially originating from interactions with electrons, $u$-quarks, and $d$-quarks. Recent systematic investigations indicate the feasibility of obtaining substantial values for $\varepsilon_{\tau \tau}$ in various radiative neutrino mass models \cite{Babu:2019mfe}. It is important to note that the approach to $\nu_\mu$ NSI is analogous and yields numerically equivalent results.

The effective strength of the NSI parameter $\varepsilon^{\text{eff}}_{\tau \tau}$ can be expressed in relation to interactions involving electrons, $u$-quarks, and $d$-quarks as follows:
\begin{equation}
    \varepsilon^{\text{eff}}_{\tau \tau} = \varepsilon_{\tau \tau}^e + (2 \varepsilon_{\tau \tau}^u + \varepsilon_{\tau \tau}^d) \frac{n_p}{n_e} + (\varepsilon_{\tau \tau}^u+ 2\varepsilon_{\tau \tau}^d) \frac{n_n}{n_e}.
\end{equation}

In various models of neutrino mass, new interactions emerge, giving rise to diverse NSI scenarios where $\varepsilon_{\tau \tau}^u$, $\varepsilon_{\tau \tau}^d$, or $\varepsilon_{\tau \tau}^e$ may be exclusively generated, or in certain instances, combinations of two NSIs or all three NSIs. For instance, a scalar leptoquark with Standard Model gauge quantum numbers ($3,2,1/6$), denoted as $\tilde{R_2}$, can induce only $\varepsilon_{\tau \tau}^d$, while the leptoquark ${R_2}$ with charges ($3,2,7/6$) exclusively leads to $\varepsilon_{\tau \tau}^u$. On the other hand, the leptoquark ${S_3}$ with charges ($\bar{3},3,1/3$) can give rise to both $\varepsilon_{\tau \tau}^d$ and $\varepsilon_{\tau \tau}^u$. Further elaboration on these scenarios can be found in the detailed discussion provided in reference \cite{Babu:2019mfe}.
To optimize the signal, we exclusively consider quark NSIs here. Although leptonic NSIs may have some impact, they are expected to be less significant compared to quark NSIs, and therefore, they do not affect the phenomenology we are concentrating on. If only $\varepsilon_{\tau \tau}^d$ exhibits a non-zero value, and under the condition of charge neutrality ($n_p=n_e$), we obtain:
%In case only $\varepsilon_{\tau \tau}^d$ is non-zero and using charge neutrality ($n_p=n_e$) we have,
\begin{equation} \label{d-nsi}
    \varepsilon^{\text{eff}}_{\tau \tau} = \varepsilon_{\tau \tau}^d \left( \frac{2-Y_e}{Y_e} \right).
\end{equation} 
Considering non-zero values of $\varepsilon_{\tau \tau}^u$, we obtain:
\begin{equation} \label{u-nsi}
    \varepsilon^{\text{eff}}_{\tau \tau} = \varepsilon_{\tau \tau}^u \left( \frac{1+Y_e}{Y_e} \right).
\end{equation}
Now, we focus on the region of the SN matter profile where $H-$ and $ L-$ resonance occur, specifically for $r > 10^4$ Km. In this particular zone, $Y_e=0.5$ (see Fig.~\ref{profile} in the Appendix). The existence of either $\varepsilon_{\tau \tau}^d$ or $\varepsilon_{\tau \tau}^u>0.33$ indicates that the effective matter potential for $\nu_\tau$ surpasses that of $\nu_e$, thereby modifying the dynamics of flavor conversion. The Hamiltonian in the presence of NSI is expressed as follows:
\begin{equation} \label{H-NSI}
    \mathcal{H}^{NSI}=\frac{1}{2 E} U\left(\begin{array}{ccc}
0 & 0 & 0 \\
0 & \Delta m_{21}^2 & 0 \\
0 & 0 & \Delta m_{31}^2
\end{array}\right) U^{\dagger}+V_{e}\left(\begin{array}{ccc}
1 & 0 & 0 \\
0 & 0 & 0 \\
0 & 0 & 3 \varepsilon^{u,d}_{\tau \tau}
\end{array}\right).
\end{equation}
For $\varepsilon^{u,d}_{\tau \tau}>0.33$, $\nu_\tau$ is the heaviest eigenstate in matter, $\nu_\tau \approx \nu_3^m(\nu_2^m)$ for NO (IO). Hence, the main difference from the standard case is that $\nu_e$ starts as the second heaviest eigenstate, $\nu_e \approx \nu_2^m (\nu_1^m)$ for NO (IO), see Fig.~\ref{E-levels-NSI}. For NO, $\nu_e$ crosses level once through a $L-$resonance, $\nu_2^m \rightarrow \nu_1^m$, at densities given by Eq.~\ref{L-res}. While for IO, $\nu_e$ is already produced as $\nu_1^m$ and reaches vacuum mostly as $\nu_1$ without crossing levels. For antineutrinos, $\bar{\nu}_e \approx \bar{\nu}_2^m(\bar{\nu}_1^m)$, and crosses levels only for NO, via $H-$resonance. In this case, however, the potential of $\nu_\tau$, $V_\tau=\varepsilon_{\tau \tau}V_e$, is higher than $V_e$, so the difference $V_e-V_\tau$ is negative. The resonance condition is found by modifying Eq.~\ref{H-res} with $V_e \rightarrow (1-\varepsilon_{\tau \tau}) V_e$.  
Neutrinos fail to satisfy this new condition in NO. However, antineutrinos in NO can fulfill this condition since their potentials exhibit an inverted sign, denoted as $\bar{V_e}=-V_e$. Furthermore, the eigenstates are not represented in the primed basis, as the potential for $\nu_\tau$ significantly surpasses the mixing terms, distinctly separating $\nu_\mu$ and $\nu_\tau$.

Assuming perfect adiabaticity of $H-$ and $L-$ resonances, we find the final flux for both orderings:
\begin{equation} \label{FeNO-NSI}
    F_e^{N O}(\varepsilon_{\tau \tau}^{u,d}>0.33) = |U_{e2}|^2 F_{ e}^i+(1-|U_{e2}|^2) F_{ x}^i \approx 0.3 F_{ e}^i+0.7 F_{ x}^i,
\end{equation}
and,
\begin{equation} \label{FeIO-NSI}
    F_e^{I O}(\varepsilon_{\tau \tau}^{u,d}>0.33) = |U_{e1}|^2 F_{ e}^i+(1-|U_{e1}|^2) F_{ x}^i \approx 0.7 F_{ e}^i+0.3 F_{ x}^i.
\end{equation}
It is noteworthy that Eq.\ref{FeNO-NSI} is equivalent to Eq.\ref{FeIO} since, in both instances, $\nu_e$ is generated as $\nu_2^m$. Consequently, the neutronization burst phase in the NO scenario with NSI and the standard IO scenario without NSI are indistinguishable. 
In the next sections, we analyze the signals from Eq.~\ref{FeNO-NSI} and Eq.~\ref{FeIO-NSI} at DUNE, emphasizing potential confusion in discerning between the two mass orderings in the presence of NSI.
%%%%%%%%%%%%%%%%%%%%%%%%%%%%%%%%%%%%%%%%%%%
%%%%%%%%%%%%%%%%%%%%%%%%%%%%%%%%%%%%%%%%%%%

\vspace{0.1 in}

%%%%%%%%%%%%%%%%%%%%%%%%%%%%%%%%%%%%%%%%%%%
%%%%%%%%%%%%%%%%%%%%%%%%%%%%%%%%%%%%%%%%%%%
\textbf{\emph{New Resonances triggered by NSI}.--}\label{sec-I-res}
%%%%%%%%%%%%%%%%%%%%%%%%%%%%%%%%%%%%%%%%%%%
%%%%%%%%%%%%%%%%%%%%%%%%%%%%%%%%%%%%%%%%%%%
\begin{figure*}[htb!]
\centering
  \includegraphics[width=0.9\textwidth]{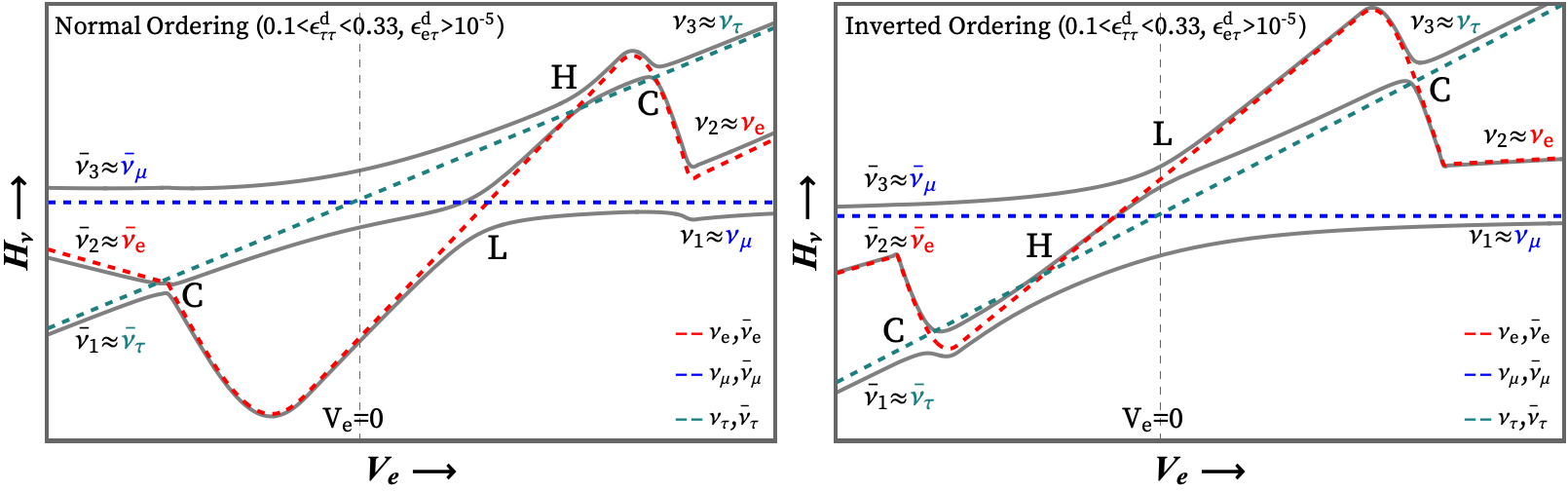}
  \caption{Configuration of energy levels for neutrinos and antineutrinos for NO (left) and IO (right) with the inclusion of $0.1<\varepsilon^d_{\tau \tau}<0.33$ and $\varepsilon^d_{e \tau}>10^{-5}$.The MSW resonance locations are marked by the letters $L$ and $H$, while the NSI-triggered resonance location is marked by the letter $C$.}
\label{E-levels-small-nsi}
\end{figure*}
We will now explore scenarios in which NSI can induce an additional resonance, designated as the $C$-resonance, alongside the existing $L-$ and $H-$ resonances.  While examining NSI within the range $0.1 < \varepsilon_{\tau \tau}^{u,d} < 0.33$, we find that it influences the flavor evolution of neutrinos during the neutronization burst phase. Despite such values of $\varepsilon_{\tau \tau}^{u,d}$ not modifying the energy level pattern in the region $r > 10^3$ Km where $L-$ and $H-$resonances occur ($Y_e \approx 0.5$),  it does alter the energy levels inside the iron core ($r < 10^3$ Km) where neutrinos are produced ($0.2 < Y_e < 0.5$). In this configuration, the $C-$ resonance emerges inside the iron core, as illustrated in Fig.~\ref{E-levels-small-nsi}. 

\begin{figure*}[htb!]
\centering
  \includegraphics[width=1\textwidth]{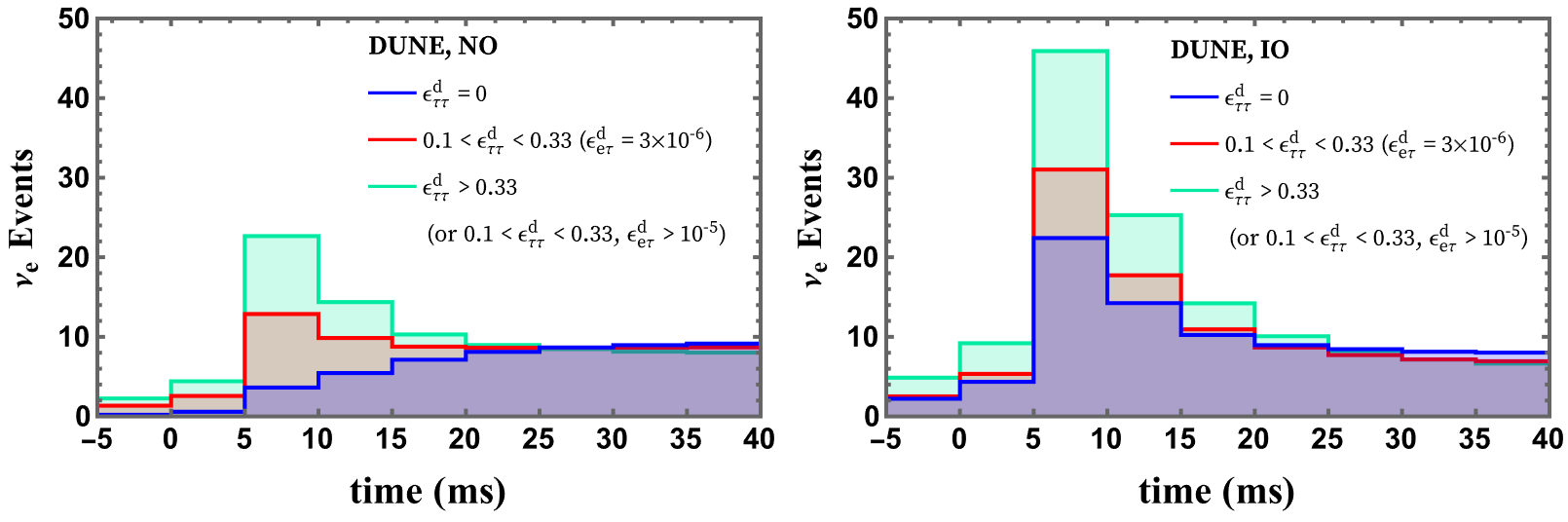}~~
  \caption{Expected number of $\nu_e$ events per bin of $5$ ms at DUNE in the time interval from $-5$ to $40$ ms corresponding to the neutronization burst phase for NO (left) and IO (right). $0$ ms represents the time of core bounce. Blue lines are computed from fluxes in Eqs.~\ref{FeNO} and \ref{FeIO}, assuming the distance between the SN and Earth to be $10$ Kpc. 
  Green lines are computed from fluxes modified by the presence of $\varepsilon_{\tau \tau}^{d}>0.33$ in Eqs.~\ref{FeNO-NSI} and \ref{FeIO-NSI}. The similarity between the green line on the left panel and the blue line on the right panel displays the degeneracy between the NO$+$NSI and standard IO cases.
  Red lines are computed from fluxes modified by the presence of $0.1<\varepsilon_{\tau \tau}<0.33$ and $10^{-6}<\varepsilon_{e \tau}<10^{-5}$. For $\varepsilon_{e \tau}$ in such interval, counting rates are intermediate between the NSI case with $\varepsilon_{\tau \tau}>0.33$ and the standard scenario. For the totally adiabatic $C-$resonance ($\varepsilon_{e \tau}>10^{-5}$), results are degenerate with the ones for $\varepsilon_{\tau \tau}>0.33$.}
\label{DUNE-spectra}
\end{figure*}

In order to examine the emergence of the $C-$resonance, we rewrite Eq.~\ref{H-NSI} by incorporating the expression for $\varepsilon_{\tau \tau}$ from Eq.~\ref{d-nsi} for $d$-quark NSI. Here, we concentrate exclusively on the matter potential term, which dominates over the vacuum term in regions of high core densities:
\begin{equation} \label{H-small-NSI}
    \mathcal{H}^{NSI} \approx V_{e}\left(\begin{array}{ccc}
1 & 0 & 0 \\
0 & 0 & 0 \\
0 & 0 & \varepsilon_{\tau \tau}^d \left( \frac{2-Y_e}{Y_e} \right)
\end{array}\right).
\end{equation}
Note that our focus here is on $d$-quark NSI. Nevertheless, the qualitative outcomes remain similar for $u$-quark NSI due to the resemblance between Eq.~\ref{d-nsi} and Eq.~\ref{u-nsi}. In the subsequent section, we analyze the consequential effects by concurrently considering NSIs involving both $d$-quark and $u$-quark.
Using the $Y_e$ values from Fig.~\ref{profile} (See Appendix for details), a level crossing occurs within the range of $50$ to $1000$ Km for $0.1<\varepsilon^d_{\tau \tau}<0.33$ as $Y_e$ transitions from $0.2$ to $0.5$. Prior to the crossing point, $\varepsilon_{\tau \tau}^d (2-Y_e)/Y_e > 1$, rendering $\nu_\tau$ as the heaviest state, and $\nu_e$ as the second heaviest: $\nu_e \approx \nu_2^m(\nu_1^m)$ for NO (IO). Subsequent to the crossing point, the scenario inverts with $\varepsilon_{\tau \tau}^d (2-Y_e)/Y_e < 1$, causing the energy levels of $\nu_e$ and $\nu_\tau$ to intersect, as depicted in Fig.~\ref{E-levels-small-nsi}.
This crossing can result in resonant conversion if we introduce off-diagonal NSI\footnote{Significant flavor off-diagonal NSI can resolve discrepancies in the determination of the standard CP-phase $\delta_{CP}$ between the NO$\nu$A and T2K long-baseline accelerator experiments \cite{Chatterjee:2020kkm, Denton:2020uda}.} $\varepsilon_{e \tau}$ and couple $\nu_e$ and $\nu_\tau$ states:
\begin{equation} \label{H-small-NSI-coupled}
    \mathcal{H}^{NSI} \approx V_{e}\left(\begin{array}{ccc}
1 & 0 & \varepsilon_{e \tau}^d \left( \frac{2-Y_e}{Y_e} \right) \\
0 & 0 & 0 \\
\varepsilon_{e \tau}^d \left( \frac{2-Y_e}{Y_e} \right) & 0 & \varepsilon_{\tau \tau}^d \left( \frac{2-Y_e}{Y_e} \right)
\end{array}\right).
\end{equation}
In such a scenario, the resonance condition can be expressed as
\begin{equation} \label{I-res}
    \varepsilon_{\tau \tau}^d \left( \frac{2-Y_e}{Y_e} \right)=1,
\end{equation}
and can be satisfied by both neutrinos and antineutrinos simultaneously. Furthermore, Eq.~\ref{I-res} is independent of matter density, energy, and mass orderings.
The adiabaticity condition is now formulated and can be expressed as
\begin{equation} \label{gamma-I}
    \gamma_C=\left| \frac{4 \left( \mathcal{H}_{e \tau}^{NSI} \right)^2}{\dot{\mathcal{H}}_{\tau \tau}^{NSI} - \dot{\mathcal{H}}_{e e}^{NSI}} \right| \approx \left| \frac{16 V_e (\varepsilon_{e\tau}^{d})^2}{Y_e \Dot{Y_e} (1+\varepsilon_{\tau \tau}^d)^3} \right| > 1.
\end{equation}
Eq.~\ref{gamma-I} is satisfied for very small values of off-diagonal NSI: $\varepsilon_{e \tau}> 10^{-5}$. The resonance is partially adiabatic for $\varepsilon_{e \tau}$ values within the range of $10^{-6}$ to $10^{-5}$. In the limit of total adiabaticity, the produced $\nu_e$ reaches vacuum as $\nu_2(\nu_1)$ for NO (IO). The scenario is analogous to the one stated in the last section by Eq.~\ref{FeNO-NSI} and Eq.~\ref{FeIO-NSI}. When adiabaticity is totally violated, the results are identical to the standard case in Eq.~\ref{FeNO} and Eq.~\ref{FeIO}. 
Partial adiabaticity provides fluxes, which are intermediary between the standard scenario and the $\varepsilon^d_{\tau \tau}>0.33$ scenario. In the next section, we analyze these signals in DUNE.

%{\color{blue}Partial adiabaticity provides fluxes which are intermediary between the standard scenario and the $\varepsilon^d_{\tau \tau}>0.33$ scenario. In the next section, we project all these signals in DUNE.}
%; see red lines in Fig.~\ref{DUNE-spectra}.
%The $\chi^2$ estimation for the case of total adiabaticity of $C-$resonance yields identical results to those in Eq.~\ref{chi-NO} and Eq.~\ref{chi-IO}: $\chi^2(NO) \approx 27$ and $\chi^2(NO) \approx 11$, respectively. 

\begin{figure}[htb!]
\centering
  \includegraphics[width=0.5\textwidth]{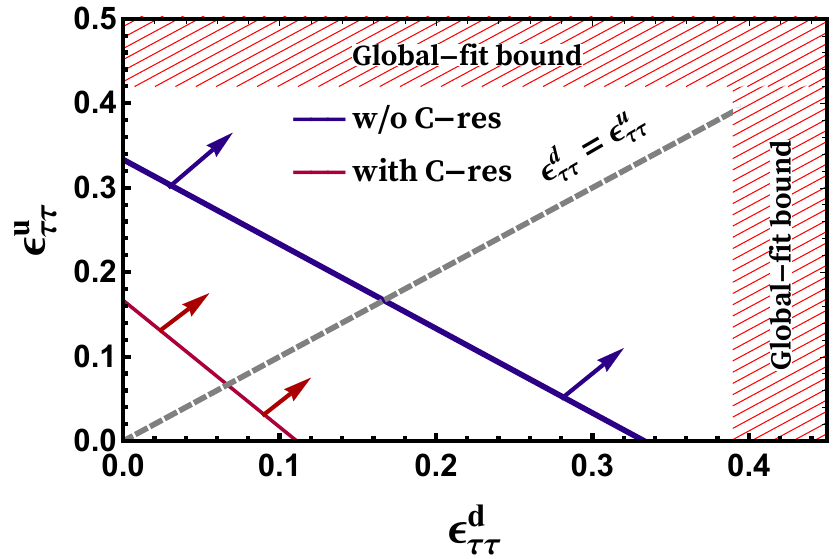}~~
  \caption{Projected sensitivities for NSI $\varepsilon_{\tau \tau}$  at DUNE. For the totally adiabatic $C-$resonance, we consider $\varepsilon_{e \tau}>10^{-5}$. Global-fit bound on NSI, as shown by the hatched region, is adopted from Ref.~\cite{Coloma:2023ixt}.}
\label{summary}
\end{figure}

We want to stress here that the search for $\varepsilon_{\tau \tau}^{u,d}$ during the neutronization burst phase is limited to values of $\mathcal{O}(0.1)$. Conversely, in later phases of SN neutrino emission, when $Y_e$ can attain values below $\mathcal{O}(0.1)$, smaller values of $\varepsilon_{\tau \tau}^{u,d}$ could potentially alter observables. Note that for $Y_e \ll 1$, Eq.~\ref{d-nsi} and Eq.~\ref{u-nsi} transforms as:
\begin{equation}
    \varepsilon_{\tau \tau}^{eff} \approx 2 \frac{\varepsilon_{\tau \tau}^d}{Y_e} \hspace{0.5 cm} \text{and} \hspace{0.5 cm} \varepsilon_{\tau \tau}^{eff} \approx  \frac{\varepsilon_{\tau \tau}^u}{Y_e}.
\end{equation}
Consequently, $\varepsilon_{\tau \tau}^{u,d}$ can be as small as $Y_e \sim O(0.01)$, while maintaining the  $\varepsilon^{eff}_{\tau \tau}$ value of order one necessary to satisfy the resonance criterion in Eq.~\ref{I-res}. Therefore, sensitivity to $\varepsilon^{u,d}_{\tau \tau}$ can be improved to $\mathcal{O}(0.01)$ by further investigating flavor conversion that occurs during other phases of neutrino emission.
%%%%%%%%%%%%%%%%%%%%%%%%%%%%%%%%%%%%%%%%%%%
%%%%%%%%%%%%%%%%%%%%%%%%%%%%%%%%%%%%%%%%%%%
\vspace{0.1 in}

%%%%%%%%%%%%%%%%%%%%%%%%%%%%%%%%%%%%%%%%%%%
%%%%%%%%%%%%%%%%%%%%%%%%%%%%%%%%%%%%%%%%%%%
{\textbf {\textit {Signal analysis at DUNE.--}}} Here, we investigate the expected SN neutrino signal spectrum at DUNE under both the standard scenario and the consideration of NSI effects. The graphical representation of the analysis, depicted in Fig.~\ref{DUNE-spectra}, illustrates the projected number of $\nu_e$ events per 5 ms interval within the time range of -5 to 40 ms (with 0 ms denoting the moment of core bounce). The scenarios for Normal Ordering (NO) and Inverted Ordering (IO) are presented on the left and right sides, respectively, assuming a distance of 10 Kpc between the supernova (SN) and Earth. For detailed information regarding the computation of DUNE event spectra, see the Appendix. The initial neutrino fluxes ($F_e^i$ and $F_x^i$) are taken from a simulation of a 15 solar mass progenitor \cite{garching}, as outlined in \cite{deGouvea:2019goq, Jana:2022tsa}. The standard case, represented by the blue lines in Fig.~\ref{DUNE-spectra} [cf. Eq.~\ref{FeNO} and Eq.~\ref{FeIO}], highlights a notable feature—the visibility of the neutronization peak, which distinguishes between the two mass orderings. The green lines are computed from Eq.~\ref{FeNO-NSI} and Eq.~\ref{FeIO-NSI} which includes NSI ($\varepsilon_{\tau \tau}^d > 0.33$). Crucially, as evident from Eq.~\ref{FeIO} and Eq.~\ref{FeNO-NSI}, along with Fig.~\ref{DUNE-spectra}, the introduction of NSI leads to a scenario where perfect confusion arises between the two mass orderings (green lines on the left panel and blue lines on the right panel are identical).

DUNE can distinguish between the standard case and signals characterized by $\varepsilon^d_{\tau \tau}>0.33$, as depicted in each panel of Fig.~\ref{DUNE-spectra}.  We analyze it using the $\chi^2$ estimator,
\begin{equation}
    \chi^2=\min _{\xi} \left( \sum_{i=1}^n 2\left[(1+\xi) F_i-D_i+D_i \ln \left(\frac{D_i}{(1+\xi) F_i}\right)\right] \right).
\end{equation}
$F_i$ and $D_i$ are the number of $\nu_e$ events in the $i$-th time bin for $\varepsilon_{\tau \tau}^d>0.33$ and $\varepsilon_{\tau \tau}^d=0$, respectively. The parameter $\xi$ is allowed to vary in the range $-1<\xi \leq 1000$ with no penalty and we select the minimum to make our derived sensitivity as much as possible independent of the overall normalization. We find for NO,
\begin{equation} \label{chi-NO}
    \chi^2(NO) \approx 27,
\end{equation}
and, for IO,
\begin{equation} \label{chi-IO}
    \chi^2(IO) \approx 11.
\end{equation}
Notably, $\chi^2(NO)$ surpasses $\chi^2(IO)$ by a considerable margin. This discrepancy is primarily attributed to the more pronounced differentiation between the green and blue lines in Fig.~\ref{DUNE-spectra} for NO, arising from the presence (green) or absence (blue) of the peak around $\sim 5$ ms. Conversely, for IO, both lines incorporate the peak, and the blue line could mimic the green one even without NSI, contingent on different spectral parameters of the supernova.
We want to stress that the green lines in Fig.~\ref{DUNE-spectra} also represent the case of NSI in the range $0.1< \varepsilon_{\tau \tau}^d < 0.33$ with adiabatic $C$-resonance ($\varepsilon_{e \tau}^d > 10^{-5}$). Partial adiabaticity of the $C$-resonance ($10^{-6} < \varepsilon_{e \tau}^d < 10^{-5}$) provides counting rates between the standard scenario and the $\varepsilon^d_{\tau \tau}>0.33$ scenario; see red lines in Fig.~\ref{DUNE-spectra}. 
We exclusively analyze scenarios with $d$-quark NSI. However, for completeness, Fig.~\ref{summary} displays projected sensitivities considering the presence of both $d$-quark and $u$-quark NSIs.

NSI affecting the flavor evolution of SN neutrinos can also be studied in the electron antineutrino ($\bar{\nu}_e$) channel using other future experiments such as Hyper-Kanmiokande \cite{Hyper-Kamiokande:2021frf} and JUNO \cite{JUNO:2023dnp}. Although these experiments will certainly provide more statistics than DUNE, their observations will lack some specific features of the time signal that are present in the $\nu_e$ channel (such as the peak at $5$ ms) that can be crucial in diagnosing the presence of NSI. The quark NSI we discussed, which is capable of inducing resonant flavor conversion of supernova neutrinos, may also be subject to complementary tests. For instance, it could produce a Glashow-like resonance feature detectable by neutrino telescopes \cite{Babu:2022fje, Babu:2019vff, Huang:2021mki} or be examined in future collider experiments, potentially resolving existing degeneracies \cite{Babu:2020nna}.
%%%%%%%%%%%%%%%%%%%%%%%%%%%%%%%%%%%%%%%%%%%
%%%%%%%%%%%%%%%%%%%%%%%%%%%%%%%%%%%%%%%%%%%

\vspace{0.1 in}

%%%%%%%%%%%%%%%%%%%%%%%%%%%%%%%%%%%%%%%%%%%
%%%%%%%%%%%%%%%%%%%%%%%%%%%%%%%%%%%%%%%%%%%
{\textbf {\textit {Conclusions.--}}}
%%%%%%%%%%%%%%%%%%%%%%%%%%%%%%%%%%%%%%%%%%%
%%%%%%%%%%%%%%%%%%%%%%%%%%%%%%%%%%%%%%%%%%%
We have shown that the NSIs have a profound impact on supernova neutrino flavor conversion, significantly impacting the neutronization burst signals. The presence of NSI can invert the energy levels of neutrino matter eigenstates and even induce a
new resonance in the inner parts close to the proto-neutron star. We showcase how the forthcoming experiments, such as DUNE, can exploit these altered energy level configurations to achieve sensitivity to NSIs at the order of $\mathcal{O}(0.1)$. Furthermore, we elaborate on how this phenomenon may lead to a puzzling ambiguity between normal and inverted mass orderings, and we thoroughly analyze its potential implications.

\vspace{0.1in}
%%%%%%%%%%%%%%%%%%%%%%%%%%%%%%%%%%%%%%%%%%%
{\textbf {\textit {Acknowledgments.--}}}
 %%%%%%%%%%%%%%%%%%%%%%%%%%%%%%%%%%%%%%%%%%%
We are grateful to Andr\'e de Gouv\^ea for an illuminating discussion and email correspondence. We also thank Pedro A. N. Machado and Manibrata Sen for useful discussions. We wish to acknowledge the Center for Theoretical Underground Physics and Related Areas (CETUP*) and the Institute for Underground Science at SURF for hospitality and for providing a stimulating environment. The work of YP was supported by the S\~{a}o Paulo Research Foundation (FAPESP) Grant No. 2023/10734-3 and 2023/01467-1 and by the National Council for Scientific and Technological Development (CNPq) Grant No. 151168/2023-7. S.J. thanks Centro de Ciências Naturais e Humanas at Universidade Federal do ABC and ICTP South American Institute for Fundamental Research for their warm hospitality during the completion of this work.
\appendix
\section*{Appendices}
 %%%%%%%%%%%%%%%%%%%%%%%%%%%%%%%%%%%%%%%%%%%
\subsection{Neutronization burst phase}\label{neutronization}
 %%%%%%%%%%%%%%%%%%%%%%%%%%%%%%%%%%%%%%%%%%%
  %%%%%%%%%%%%%%%%%%%%%%%%%%%%%%%%%%%%%%%%%%%
\begin{figure}[htb!]
\centering
  \includegraphics[width=0.5\textwidth]{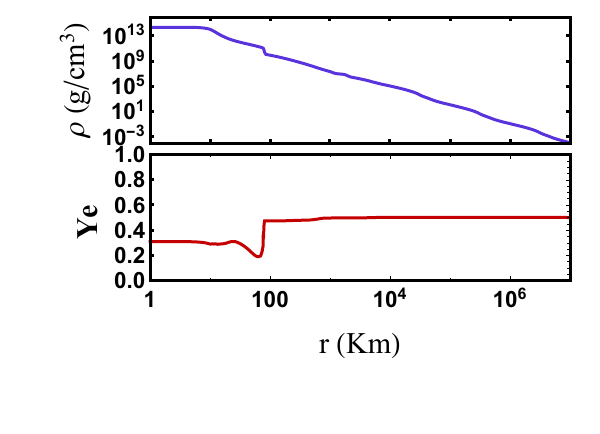}
  \caption{Status of the matter density $\rho$ and and electron number fraction $Y_e$ at $4.37$ ms after core bounce. \cite{Serpico:2011ir,Tang:2020pkp}.}
\label{profile}
\end{figure}
 %%%%%%%%%%%%%%%%%%%%%%%%%%%%%%%%%%%%%%%%%%%
The SN neutronization burst phase encompasses the period of $\sim 40$ ms after core bounce when the $\nu_e$ luminosity dominates over all other flavors, $\bar{\nu}_e$, $\nu_x$ and $\bar{\nu}_x$, with $x=\mu,\tau$ ($\nu_e$ luminosity can be $10-100$ times higher than luminosity in other flavors, see \cite{deGouvea:2019goq,Jana:2022tsa}). During the core collapse, the inner core is compressed and reaches nuclear densities, while the matter falling above it bounces back and launches a shock wave, which dissociates nuclei into its component nucleons as it travels outwards. The capture of electrons in the environment by the dissociated protons, $e^{-}+p \rightarrow n +\nu_e$, is responsible for generating the large $\nu_e$ burst during the neutronization burst phase. Theoretical uncertainties in the calculation of the neutronization fluxes are believed to be as small and $10 \%$ \cite{Serpico:2011ir, Wallace:2015xma, OConnor:2018sti, Kachelriess:2004ds} and uncertainties related to neutrino-neutrino refraction \cite{Mirizzi:2015eza, Capanema:2024hdm} can also be avoided. Therefore, the burst phase presents a great opportunity to search for new physics with SN neutrinos.

Neutrinos are produced in the region $r < 100$ Km, which is opaque due to very high densities ($\rho > 10^{11}$ $\text{g/cm}^3$), and, at $r\sim 100$ Km, they start free streaming. The efficient electron capture at the production region reduces its electron number fraction, $Y_e=n_e/(n_p+n_n)$, where $n_e$, $n_p$ and $n_n$ are the electron, proton, and neutron number densities, respectively, to levels below the one found in the envelope ($Y_e=0.5$). Precisely, this difference between $Y_e$ in inner and outer layers might produce observable consequences for non-zero NSI. Further details can be seen in Fig.~\ref{profile}, where we plot $\rho$ and $Y_e$ at $\sim 5$ ms \cite{Fischer:2009af,Tang:2020pkp}. This instant is representative of the matter profile during the neutronization peak, the most distinct feature of the burst phase.
\vspace{0.1in}
%%%%%%%%%%%%%%%%%%%%%%%%%%%%%%%%%%%%%%%%%%%
\subsection{DUNE: technical details}
%%%%%%%%%%%%%%%%%%%%%%%%%%%%%%%%%%%%%%%%%%%
The Deep Underground Neutrino Experiment (DUNE) will be composed of four-time projection chambers, each with $10$ Kton of liquid argon, placed underground in the Long-Baseline Neutrino Facility (LNBF) in South Dakota, United States \cite{DUNE:2020zfm, Cuesta:2023nnt}. DUNE will primarily detect neutrinos with energies of GeV and higher coming from a beam produced at Fermilab. Nevertheless, DUNE is also sensitive to SN neutrinos in the range from about $5$ MeV to tens of MeV via the charged-current interaction
\begin{equation} \label{ArCC}
   \nu_e+{ }^{40} \mathrm{Ar} \rightarrow{ }^{40} \mathrm{~K}^*+e^{-},
\end{equation}
that separates the electron flavor and has a much higher cross-section than the elastic scattering, $\nu_{e, \mu, \tau}+e^- \rightarrow \nu_{e, \mu, \tau} + e^-$, which is the current main detection channel for $\nu_e$ \cite{Scholberg:2012id, Zhu:2018rwc}.
We compute the event spectrum for $\nu_e$ at DUNE as \cite{Capozzi:2018rzl}
\begin{equation}
    \frac{d N_{\nu_e}}{d E_{r}}=\frac{N_{\mathrm{Ar}}}{4 \pi R^2} \int d E_{\nu_e} F_{\nu_e}\left(E_{\nu_e} \right) \sigma_{\nu_e+\mathrm{Ar}}\left(E_{\nu_e} \right) W\left(E_{r}, E_{\nu_e} \right)
\end{equation}
where $N_{\mathrm{Ar}}$ is the number of target $^{40}\mathrm{Ar}$ nuclei in the chambers, $R$ is the distance between the SN and the Earth, $F_{\nu_e}$ is the $\nu_e$ flux leaving the SN with energy $E_{\nu_e}$, $\sigma_{\nu_e+\mathrm{Ar}}$ is the cross section for the interaction in Eq.~\ref{ArCC} generated by MARLEY \cite{Gardiner:2018zfg}, $W$ is the gaussian energy resolution with $\sigma_E/\text{MeV}=0.11 \sqrt{E_{r}/\text{MeV}}+0.02 E_{r}/\text{MeV}$ and $E_{r}$ is the reconstructed electron energy.
%%%%%%%%%%%%%%%%%%%%%%%%%%%%%%%%%%%%%%%%%%%
\bibliographystyle{utcaps_mod}
\bibliography{reference}
\end{document}